\documentclass[aps,pra,twocolumn,superscriptaddress,nobibnotes,letterpaper]{revtex4}
\usepackage{times}
\usepackage[pdftex]{graphicx}
\usepackage[pdftex]{epsfig}
\usepackage{amsmath}
\usepackage{pslatex}
\usepackage{amssymb}
\usepackage{amsfonts}
\usepackage{color}
\usepackage{wrapfig}
\usepackage{eucal}
\usepackage{hhline}
\usepackage{threeparttable}
\usepackage{supertabular}
\usepackage{multirow}
\usepackage{tabularx}
\usepackage{setspace}
\usepackage[compact]{titlesec}
\usepackage{hyperref}
\usepackage{graphicx}
\usepackage{esint}
\usepackage{verbatim}
\usepackage{units}

\newcommand{\nbar}{\langle n \rangle}
\newcommand{\nbarmin}{\langle n_{\text{min}} \rangle}
\newcommand{\nbath}{\bar{n}_{\text{b}}}

\newcommand{\Tbath}{T_{\text{b}}}

\newcommand{\Tm}{T_{\text{m}}}
\newcommand{\Gammath}{\Gamma_{\text{th}}}
\newcommand{\Gammadeph}{\Gamma_{\phi}}
\newcommand{\Gammameas}{\Gamma_{\text{meas}}}
\newcommand{\Hint}{\hat{H}_{\text{int}}}
\newcommand{\ncavOop}{\hat{n}_{\text{c}}}
\newcommand{\Frp}{F_{\text{rp}}}

\newcommand{\Fth}{F_{\text{th}}}
\newcommand{\SiN}{\text{Si}_3\text{N}_4}
\newcommand{\meff}{m_{\text{eff}}}
\newcommand{\Mmode}{\text{IP}_{d,0}}
\newcommand{\Omode}{\text{TE}_{e,0}}
\newcommand{\thetaH}{\theta_{h}}
\newcommand{\thetafb}{\theta_{\text{fb}}}
\newcommand{\omegafb}{\omega_{\text{fb}}}

\newcommand{\ncavO}{n_{\text{c}}}

\newcommand{\nimp}{n_{{\text{imp}}}}
\newcommand{\nimpSN}{n_{{\text{imp,SN}}}}

\newcommand{\nBA}{n_{{\text{BA}}}}
\newcommand{\nBASN}{n_{{\text{BA,SN}}}}
\newcommand{\nBAabs}{n_{{\text{BA,abs}}}}
\newcommand{\nsql}{n_{\text{SQL}}}
\newcommand{\gzeroO}{g_{\text{0}}}
\newcommand{\gOMO}{g_{\text{OM}}}
\newcommand{\deltal}{\Delta}
\newcommand{\kappaO}{\kappa}

\newcommand{\kappae}{\kappa_{\text{e}}}

\newcommand{\gammaiO}{\gamma_{\text{i}}}

\newcommand{\omegacO}{\omega_{\text{c}}}
\newcommand{\omegam}{\omega_{\text{m}}}  
\newcommand{\omegamO}{\omega_{\text{m}}}
\newcommand{\omegamiO}{\omega_{\text{m,0}}}   
\newcommand{\omegaL}{\omega_{\text{l}}} 
\newcommand{\fmO}{f_{\text{m}}}

\newcommand{\QoO}{Q_{\text{c}}}
\newcommand{\QmO}{Q_{\text{m}}}
\newcommand{\QmOi}{Q_{\text{m,i}}}

\newcommand{\etac}{\eta_{\text{c}}}
\newcommand{\etaQE}{\eta_{\text{QE}}}
\newcommand{\etaf}{\eta_{\text{f}}}

\newcommand{\xzpfO}{x_{\text{ZPF}}}

\newcommand{\taum}{\tau_{\text{m}}}
\newcommand{\Csp}{C_1}
\newcommand{\kB}{k_{\text{B}}}
\newcommand{\xn}{x_\text{n}}
\newcommand{\xmeas}{y}

\newcommand{\xmeasdot}{\dot{y}}

\newcommand{\etaT}{\eta_\text{t}}
\newcommand{\etaR}{\eta_\text{r}}

\newcommand{\gopt}{g_{\text{opt}}}
\newcommand{\nbathhalf}{\nbath + 1/2}
\newcommand{\eff}{\text{eff}}
\newcommand{\geecd}{g_{\text{cd}}}
\newcommand{\Omodej}{\text{TE}_{e(o),j}}
\newcommand{\omegafbopt}{\omega_{\text{fb,opt}}}

\newcommand{\Figref}[1]{Fig.~\ref{#1}}
\newcommand{\Eqref}[1]{eqn.~(\ref{#1})}
\newcommand{\smalltimes}{\!\times\!}

\begin{document}

\title{Optical read out and feedback cooling of a nanostring optomechanical cavity}

\author{Alex G.\ Krause}
\thanks{These authors contributed equally to this work.}
\author{Tim D. Blasius}
\thanks{These authors contributed equally to this work.}
\author{Oskar Painter}
\email{opainter@caltech.edu}
\affiliation{Kavli Nanoscience Institute and Thomas J. Watson, Sr., Laboratory of Applied Physics, California Institute of Technology, Pasadena, CA 91125, USA}
\affiliation{Institute for Quantum Information and Matter, California Institute of Technology, Pasadena, CA 91125, USA}

\date{\today}
\begin{abstract}
Optical measurement of the motion of a $940$~kHz mechanical resonance of a silicon nitride nanostring resonator is demonstrated with a read out noise imprecision reaching $37$~dB below that of the resonator's zero-point fluctuations.  Via intensity modulation of the optical probe laser, radiation pressure feedback is used to cool and damp the mechanical mode from an initial room temperature occupancy of $\nbath = 6.5\times 10^6$ ($\Tbath=295$~K) down to a phonon occupation of $\nbar = 66 \pm 10$, representing a mode temperature of $\Tm \approx 3$~mK.  The five decades of cooling is enabled by the system's large single-photon cooperativity $(\Csp = 4)$ and high quantum efficiency of optical motion detection ($\eta_{t} = 0.27$).
\end{abstract}
\maketitle

Cavity-optomechanical systems utilize multi-pass scattering of light within a cavity to perform sensitive measurement of mechanical motion, with applications ranging from inertial microsensors~\cite{krause_high-resolution_2012,gavartin_hybrid_2012} to transducers for interfacing disparate quantum systems~\cite{andrews_bidirectional_2014,Joeckel2014}. In this work, we integrate a silicon nitride nanostring mechanical resonator of motional mass $\meff = 90$~picograms and frequency $\omegamO/2\pi=940$~kHz, with a photonic crystal optical nanocavity.  The strength of the optomechanical coupling in this structure is characterized by a per photon measurement rate of the nanostring motion which is four times that of its intrinsic mechanical damping rate.  Combined with an overall optical detection efficiency of $\eta_{t} = 0.27$, this enables a measurement imprecision which reaches $37$~dB below that of the zero-point fluctuation noise of the bare mechanical resonator.  Active cancellation of the mechanical thermal motion through feedback on the read-out laser's intensity realizes cooling~\cite{Mancini1998,genes_ground-state_2008} from room temperature down to a phonon occupancy of $\nbar =66 \pm 10$, corresponding to an effective mode temperature of $\Tm = 3$~mK.  This chip-scale microresonator, operating in a room temperature environment yet close to its quantum ground-state of motion, has a thermal-noise-limited force sensitivity of $125$~aN/Hz$^{1/2}$, a bandwidth of $190$~kHz around resonance, and a linear dynamic range at one second integration time of greater than $60$~dB.

Resolved-sideband radiation pressure cooling has recently been demonstrated~\cite{teufel_sideband_2011,chan_laser_2011} as an effective means to cool a targeted mechanical resonance of a structure close its quantum mechanical ground-state of motion.  This technique, sharing similar physics to the resolved-sideband cooling of trapped ions~\cite{Monroe1995}, requires spectral filtering of the upper (anti-Stokes) motional sideband from the lower (Stokes) motional sideband by a high-$Q$ cavity in which the cavity linewidth ($\kappa$) is narrower than the mechanical resonances frequency ($\omegamO$).  To date, experiments involving resolved-sideband cooling of mesoscopic mechanical objects to their quantum ground-state have relied on cryogenic pre-cooling using conventional refrigeration means.  Bath temperatures $\Tbath \lesssim 100$~mK are utilized for microwave devices~\cite{teufel_sideband_2011} to enable high-$Q$ superconducting cavities, whereas in the optical domain~\cite{chan_laser_2011} more modest bath temperatures of $\sim 10$~K in a helium cryostat have been employed to, among other things, reduce intrinsic mechanical damping.

An alternative method of radiation pressure cooling, one which is more amenable to lower frequency mechanical resonators, relies on low noise optical read out of mechanical displacement combined with active feedback of the optical probe intensity~\cite{Mancini1998,Courty2001,genes_ground-state_2008}.  Previous optomechanical feedback cooling experiments~\cite{Cohadon1999,li_millikelvin_2011,gieseler_subkelvin_2012,poggio_feedback_2007,abbott_observation_2009,DWilson_PhDthesis2012} have demonstrated the ability to cool a wide range of mechanical resonators, from suspended large scale kilogram mass mirrors~\cite{abbott_observation_2009} to optically levitated microspheres~\cite{li_millikelvin_2011,gieseler_subkelvin_2012}.  Although these experiments have realized substantial cooling, reaching the quantum mechanical ground-state of motion using active feedback cooling remains an illusive goal due to the stringent requirements on the measurement imprecision and the mechanical resonator $Q$-factor. Cooling to the ground-state requires a displacement measurement capable of resolving motion at the level of the quantum zero-point fluctuations of the mechanical resonator within its thermal decoherence time, and with back-action close to the Heisenberg limit~\cite{Clerk_RMP_2010,Bushev2006,teufel_sideband_2011,Wilson2014}.  This regime has recently been approached in several cavity-optomechanical systems at liquid helium temperatures~\cite{safavi-naeini_squeezed_2013,purdy_observation_2013,Wilson2014}, with feedback cooling of a MHz-frequency nanostring resonator being demonstrated in Ref.~\cite{Wilson2014} down to an occupancy of $5$ phonons.  Here we set out to explore limits to feedback cooling of a photonic crystal nanostring structure in a room temperature environment suitable for a broad range of sensing applications.       

\begin{figure*}[t]
\includegraphics[width=2\columnwidth]{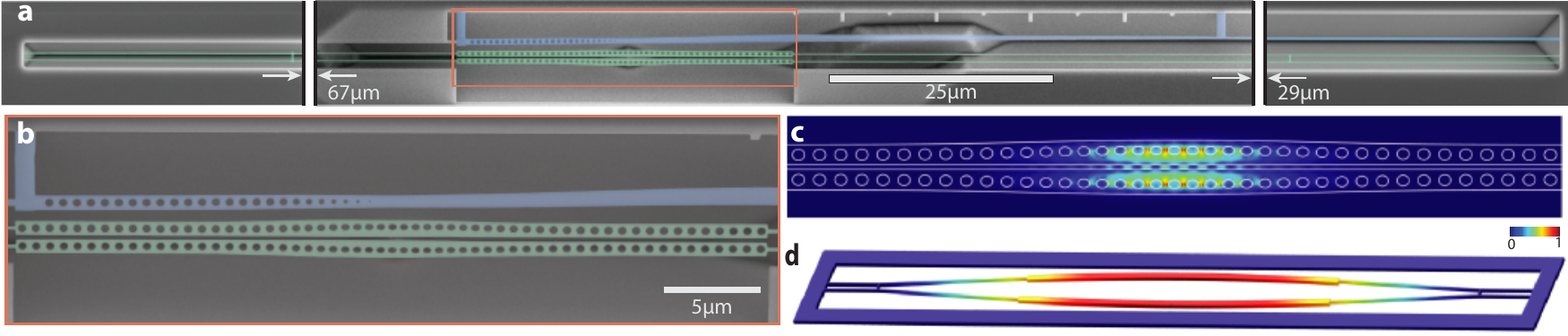}
\caption{\textbf{a}, False color SEM image of a device as used in the experiment, made from highly stressed, 415nm thick, stoichiometric silicon nitride released from a silicon wafer. The device consists of a photonic crystal cavity suspended on either side by nanoscale cross-section ($115~\mu\textrm{m}\smalltimes130$~nm) tethers. The green and blue overlay regions indicate the photonic crystal cavity and the coupling waveguide, respectively. Due to the extreme aspect ratio of this device, only the ends of the tether region are shown, with the extent of the missing gaps on either side of the center of the device indicated.  \textbf{b}, Zoomed-in SEM image of the photonic crystal section (green) and the adiabatically tapered on-chip coupling waveguide (blue). \textbf{c}, FEM-simulated electromagnetic energy density of the $\Omode$ optical resonance, with the outline of the $\SiN$ beam shown in white. \textbf{d}, FEM simulation of the first-order in-plane differential mechanical resonance ($\Mmode$) of a dual nanobeam structure, indicating the displacement of the beams.  For clarity, the tethers in the simulated structure are shorter than those of the actual device.  In \textbf{c} the color scalebar indicates large (red) and small (blue) optical energy density, whereas in \textbf{d} the scalebar indicates large (red) and small (blue) displacement amplitude.}\label{Fig_DEV}
\end{figure*}

Cooling with simple derivative feedback can be understood from the following (classical) harmonic oscillator equation of motion,
\begin{align}
\ddot{x} + \gammaiO\dot{x} + \omegamO^2x & =\frac{\Fth}{\meff} - \left(\gammaiO  g e^{i\thetafb}\right)\xmeasdot,  \label{SHO_EOM}
\end{align}
\noindent where $x(t)$ is the amplitude of motion of the mechanical resonator, $\meff$ is the motional mass of the mechanical resonator, $\omegamO$ is the mechanical resonance frequency, $\gammaiO$ the intrinsic mechanical energy decay rate of the resonator, and $\Fth$ is the effective noise force of the thermal bath coupled to the mechanical resonator.  The final term on the right-hand side of \Eqref{SHO_EOM} is the feedback forcing term, where $g$ is the unitless gain of the feedback loop, $\thetafb$ the phase of the feedback, $\xmeas (t) = x(t) + \xn (t)$ is the estimated resonator's amplitude of motion from measurement, and $\xn(t)$ is the measurement noise (error).  When $\xn(t)$ is negligible and $\thetafb = 0$, this term leads to viscous damping and cooling of the mechanical resonance, with the closed-loop mechanical $Q$-factor and phonon occupancy scaling as $\QmO = \QmOi/(1+g)$ and $\nbar = \nbath/(1+g)$, respectively.  Here, $\QmOi$ is the intrinsic mechanical quality factor, $\nbath$ is the thermal bath occupancy at $\omegamO$, and $\nbar$ is the resulting average phonon occupancy of the mechanical mode under feedback.  For the room temperature measurements of this work, $\nbath \approx \kB\Tbath/\hbar\omegamO = 6.5\times 10^{6}$, where $\Tbath = 295$~K is the bath temperature, $\kB$ is Boltzmann's constant, and $\hbar$ is Planck's constant over $2\pi$.

A more rigorous derivation of feedback cooling including quantum measurement noise~\cite{genes_ground-state_2008,Courty2001} shows that in the limit of large feedback gain ($g \gg 1$) the cooled phonon occupancy is approximately given by,

\begin{align}
\nbar + \frac{1}{2} \approx \left(\frac{\nbath + 1/2}{g}\right) + \left(g\nimp + \frac{\nBA}{g}\right).  \label{nbar_genes_asp}
\end{align}

\noindent Here $\nimp$ is the open-loop measurement imprecision in units of photon number of the undamped oscillator, and $\nBA$ is the open-loop quantum back-action noise of the optical position measurement.  The first term on the right hand side of \Eqref{nbar_genes_asp} represents the damped thermal noise from the bath, and is limited by the achievable feedback gain.  The quantum fluctuations of the probe laser light manifest as radiation pressure shot noise~\cite{purdy_observation_2013}, imposing a quantum-limited relation between the imprecision and back-action noise sources, $\nBASN = 1/(16\etaT\nimpSN)$, where $\etaT$ is the quantum efficiency of detection of light that enters into the optomechanical cavity and is scattered by the mechanical resonator.  The last two terms on the right hand side of \Eqref{nbar_genes_asp} thus represent a measurement limit to the attainable cooling. The optimal feedback gain is $\gopt = \sqrt{\left(\nbath + \nBA \right)/\nimp}$, which yields the minimum attainable resonator occupancy assuming only quantum back-action, $\left(\nbar_{\text{min}} + 1/2\right) = 2\sqrt{\nimp\nbath + 1/16\etaT}$. Achieving $\nbar<1$ thus requires both $\nimp < 1/2\nbath$ and $\etaT > 1/9$.  Consideration of the feedback bandwidth brings an additional constraint~\cite{genes_ground-state_2008}, requiring $\QmOi > \nbath$.  In addition to these fundamental constraints, technical limitations such as optical-absorption heating and thermo-refractive noise~\cite{Braginsky2000} may also play a role, adding excess back-action and imprecision, respectively.


Shown in Fig.~\ref{Fig_DEV}, the optomechanical structure studied in this work consists of a ``zipper'' photonic crystal optical cavity~\cite{eichenfield_picogram-_2009} supported by nanoscale tethers.  The structure is fabricated using standard electron beam lithography and plasma etching techniques, and formed out of a $415$~nm thick layer of stoichiometric silicon nitride ($\SiN$) deposited on a silicon (Si) handle wafer.  The zipper optical cavity (green shaded region of \Figref{Fig_DEV}b) consists of two micron-wide beams with linear hole patterning, separated by a small gap of $s=150$~nm, and attached to the bulk by $115$~$\mu$m long nanotethers of width $w=130$~nm. The optical cavity design was simulated and optimized~\cite{chan_optimized_2012} using the COMSOL finite-element-method (FEM) mode solver. The optical cavity mode of interest is the fundamental even mode ($\Omode$) with electric field polarization predominantly in the plane of the $\SiN$ film.  The central modification of the hole shape and location strongly confines the electromagnetic energy in the gap between the beams of the optical cavity (see \Figref{Fig_DEV}c), which results in a large shift in the $\Omode$ resonance frequency ($\omegacO$) with relative in-plane displacement of the beams.  The mechanical mode of interest is the fundamental in-plane differential mode of the beams ($\Mmode$) depicted in Fig.~\ref{Fig_DEV}d.  For the geometry considered here, the optical resonance wavelength is in the $1500$~nm band ($\omegacO \sim 190$~THz), and the mechanical resonance frequency is $\omegamO/2\pi=940$~kHz.  The simulated motional mass and zero-point motion amplitude of the $\Mmode$ mechanical mode are $\meff = 90\times10^{-15}$~kg and $\xzpfO = \sqrt{\hbar/(2\meff\omegamO)} = 9.7$~fm, respectively, where the generalized coordinate of mechanical motion, $x$, is chosen to be the point of maximum in-plane displacement of the beams. 

To lowest order in the mechanical amplitude, the sensitivity of the optical resonance frequency to mechanical motion is quantified by the linear dispersive coupling parameter $\gOMO \equiv \partial\omegacO/\partial x$.  For the optical $\Omode$ and the mechanical $\Mmode$ the optomechanical coupling is simulated to be $\gOMO/2\pi = 41$~GHz/nm, corresponding to a vacuum coupling rate of $\gzeroO = \gOMO\xzpfO = 2\pi[358$~kHz$]$.  The interaction Hamiltonian of the coupled cavity-optomechanical system is given by $\Hint = \hbar\gOMO\ncavOop\hat{x}$, where $\ncavOop$ ($\ncavO$) is the intra-cavity photon number operator (average).  The corresponding average radiation pressure force applied by the intra-cavity optical field on the mechanical resonator is $\Frp = -\langle\partial \Hint/\partial \hat{x}\rangle = -\hbar \gOMO\ncavO$.  By modulating the laser intensity input to the optical cavity we can create a feedback cooling force as in \Eqref{SHO_EOM}.  $\gOMO$ is therefore a critical parameter, determining both the optical measurement sensitivity to mechanical motion and the strength of the radiation pressure force that can be applied per photon.  In principle, small $\gOMO$ can be overcome with larger optical power.  In practice, parasitic effects such as thermo-refractive noise~\cite{Braginsky2000} and optical-absorption heating then tend to limit the achievable measurement imprecision and back-action~\cite{Wilson2014}.  In this regard, a figure of merit is the single-photon cooperativity ($\Csp$), which physically represents the ratio of the displacement measurement rate per photon to the decoherence rate per (thermal bath) phonon of the mechanical resonator. For the zipper cavity studied here, $\Csp=4\gzeroO^2/\kappa\gammaiO = 4$.

\begin{figure}[t]
\includegraphics[width=\columnwidth]{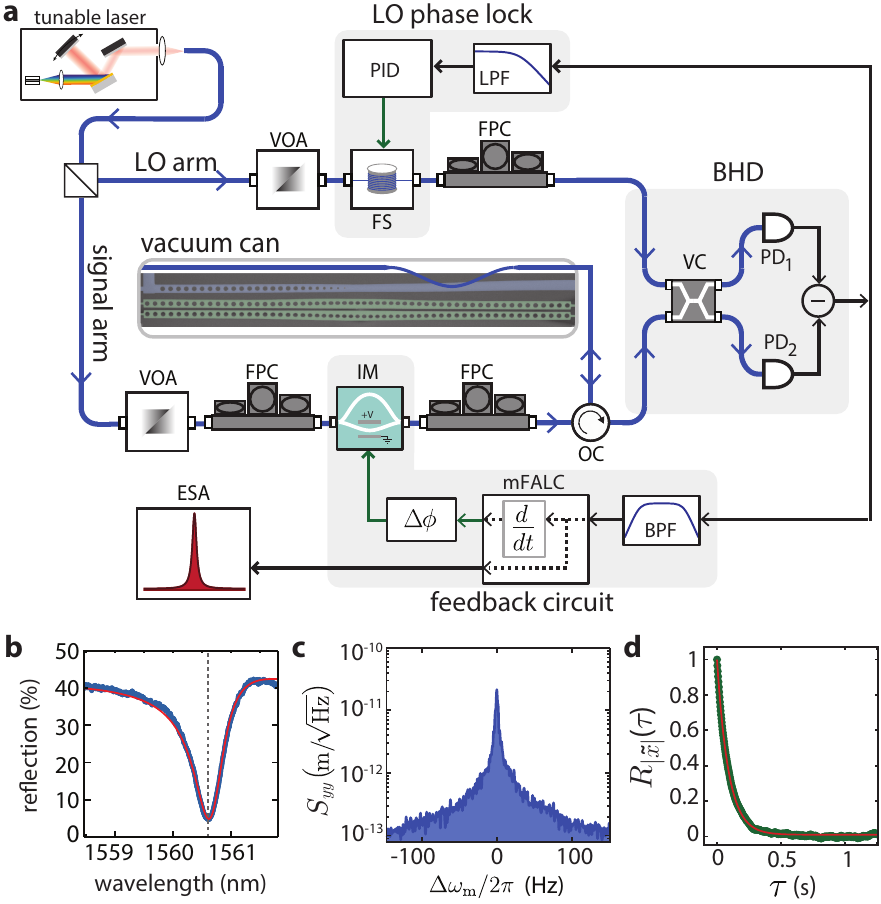}
\caption{ \textbf{Device Characterization and Experimental Setup.} \textbf{a}, Schematic of the optical and electrical set-up used to characterize and feedback-cool the mechanical resonator. Additional acronym: fiber polarization controller (FPC). \textbf{b}, Plot of the laser reflection spectrum when scanned across the optical mode used in the experiment (blue) and fit (red). Dashed grey line indicates on-resonance laser position during measurements. The measured loaded optical $Q$-factor is $\QoO = 2.5\smalltimes10^3$, with a waveguide loading to total cavity loss rate ratio of $\kappae/\kappaO=0.63$. \textbf{c}, Power spectral density (PSD) of the mechanical resonator's thermal noise near $940$~kHz, transduced using the setup in \textbf(a) with the laser on-resonance, $\deltal = 0$, and at low power $\ncavO \lesssim 1$. \textbf{d}, Autocorrelation of the slowly varying amplitude of the transduced mechanical thermal noise, $R_{|\tilde{x}|}(\tau) = \langle|\tilde{x}(t)||\tilde{x}\left(t+\tau\right)| \rangle$. $x(t) \equiv \tilde{x}(t)\sin(\omegamO t + \phi(t))$) defines the slowly varying amplitude, $\tilde{x}(t)$.  An exponential fit (red curve) to the measured data (green circles) yields an intrinsic mechanical damping rate $\gammaiO/2\pi = 1.76$~Hz and quality factor $\QmOi = 5.3 \times 10^5$.}\label{Fig_Setup}
\end{figure}

The experimental setup used to characterize the optical and mechanical properties of the zipper cavity is shown in \Figref{Fig_Setup}(a).  The device is mounted inside a vacuum chamber reaching a pressures of $2.5 \times 10^{-5}$~mbar, sufficient to eliminate the effects of gas-damping of the mechanics.  A tunable external cavity semiconductor diode laser (New Focus Velocity series) is used to provide both the signal beam on-resonance with the zipper cavity and the local oscillator beam (LO) for homodyne detection. Laser light is efficiently coupled into and out of the zipper cavity using an optical fiber taper~\cite{michael_optical_2007} in combination with an on-chip tapered waveguide (blue shaded region of Fig.~\ref{Fig_DEV}).  Tapering the width of the on-chip waveguide allows for adiabatic mode-conversion between the waveguide and the tapered fiber placed upon it as in Ref.~\cite{groblacher_highly_2013}. The reflected optical signal from the zipper cavity is separated using an optical circulator and sent to a balanced homodyne detector (BHD).  A low-pass filtered (LPF; bandwidth $<200$~kHz) version of the BHD signal is sent to a control circuit (PID) which drives a fiber stretcher (FS) to lock the relative phase ($\thetaH$) between optical LO and reflected signal beam, and sets the phase quadrature of the homodyne detected signal.  A band-pass filtered (BPF; bandwidth $0.2$-$1.9$~MHz) version of the BHD signal is sent to an electronic spectrum analyzer (ESA) to measure the mechanical noise spectrum.  In the case of optical feedback, the band-pass filtered signal is also sent to an analog differentiator circuit (Toptica mFALC), whose output is sent through a variable phase shifter, $\Delta \phi$, and finally onto an electro-optic intensity modulator (IM) which closes the feedback loop and modulates the signal beam intensity.

In Fig.~\ref{Fig_Setup}b we plot the reflected optical power normalized by the input power at the device as the laser frequency is scanned across the $\Omode$ optical resonance of the zipper cavity.  The background level of this normalized plot, $\etaR=0.43$, determines the overall detection efficiency of light emitted from the cavity, including the single-pass coupling between the chip waveguide and optical fiber taper ($\etac = 0.72$), optical loss between the fiber taper and BHD photodetectors ($\etaf=0.69$), and the quantum efficiency of the photodetectors ($\etaQE=0.88$). The measured linewidth of the optical cavity resonance is $\kappa/2\pi = 77$~GHz, corresponding to a loaded optical $Q$-factor of $\QoO = 2.5 \times 10^3$.  A low optical $Q$ was chosen to increase the optical power handling ability of the zipper cavity. From the depth and Fano-like shape of the resonance dip, the coupling rate of the cavity to the on-chip waveguide is estimated to be $\kappae/2\pi = 49$~GHz, yielding a slightly overcoupled system with $\kappae/\kappa = 0.63$.  The total efficiency of motion detection is related to the detection of laser light that enters the cavity and interacts with the mechanical resonator, which for this device and set-up is $\etaT = \etaR (\kappae/\kappa) = 0.27$.    

Measurement of the mechanical motion, $x(t)$, is performed by setting the laser frequency ($\omegaL$) to the resonance frequency ($\omegacO$) of the optical mode ($\Delta \equiv \omegaL-\omegacO = 0$), and monitoring the phase quadrature ($\thetaH = \pi/2$) on the balanced homodyne detector~\cite{safavi-naeini_squeezed_2013}.  The measured single-sided noise power spectral density (NPSD) is plotted without feedback (open loop) in Fig.~\ref{Fig_Setup}c, showing the optically transduced thermal Brownian motion of the $\Mmode$ mechanical resonance at $940$~kHz.  Here, the NPSD is calibrated using the measured $\gOMO$ of the $\Mmode$ mechanical resonance (along with optical input power, optical cavity parameters, and optical detection efficiency), and plotted as $S_{yy}(\omega)$ in units of m/$\sqrt{\text{Hz}}$.  The optomechanical coupling is inferred from the optical spring shift of the mechanical resonance frequency~\cite{eichenfield_picogram-_2009}, and is measured to be $\gOMO/2\pi = 36$~GHz/nm, in close agreement with the simulated coupling coefficient for the measured device geometry.  Autocorrelation of the magnitude of the slowly varying amplitude of the resonator thermal motion, $R_{|\tilde{x}|}(\tau) = \langle |\tilde{x}(t+\tau)||\tilde{x}(t)|\rangle$, is shown in Fig.~\ref{Fig_Setup}d.  A fit to the exponential decay of $R_{|\tilde{x}|}(\tau)$ (red curve) yields an intrinsic mechanical damping factor of $\gammaiO/2\pi=1.76$~Hz, corresponding to an intrinsic mechanical $Q$-factor of $\QmOi = 5.3 \times 10^{5}$.

\begin{figure}[t]
\includegraphics[width=\columnwidth]{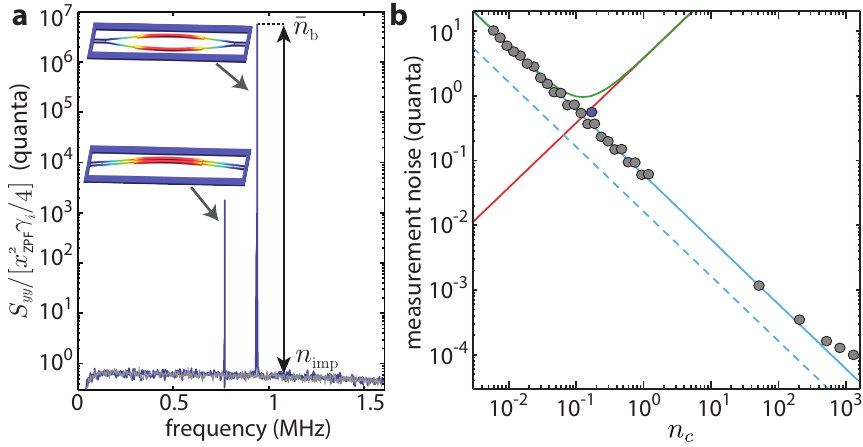}
\caption{\textbf{Measurement noise.} \textbf{a}, Measured wideband NPSD plotted in units of phonon quanta of the dominant mechanical mode at $940$~kHz for $\ncavO = 0.17$ (blue) and with the signal beam blocked ($\ncavO = 0$; grey curve). The insets of (a) are simulated mechanical mode displacement profiles for the two mechanical resonances visible in the spectrum. The mode at $740$~kHz is the poorly transduced in-plane common motion of the zipper beams. Frequencies below $200$~kHz are attenuated by a high-pass filter and the small slope of the background noise level is a result of the frequency-dependent gain of the balanced photodetectors. \textbf{b}, Plot of the measured imprecision noise in units of phonon quanta ($\nimp$; grey filled circles).  Also plotted are the theoretical vacuum-noise-limited imprecision ($\nimpSN$; solid cyan curve), theoretical quantum back-action noise ($\nBASN$; solid red curve), and theoretical quantum-limited measurement noise ($\nimpSN + \nBASN$; solid green curve), assuming the measured $\eta_T = 0.27$ and with no additional fit parameters. Dashed curve is the measurement imprecision for an ideal continuous position measurement with $\eta_T =1$.} \label{Fig_SQL}
\end{figure}

Figure~\ref{Fig_SQL}a shows an open-loop, wideband spectrum of the measured mechanical NPSD ($\Delta=0$, $\thetaH = \pi/2$) for an intracavity photon number of $\ncavO = 0.17$ (blue curve) and for $\ncavO = 0$ (grey curve).  Measurement of the noise level with the signal arm blocked corresponds very closely to the signal vacuum-noise level (electronic detector noise is $12.9$~dB below the measured noise level).  Here we have plotted the NPSD in units of phonon quanta, $S_{yy}(\omega)/(\xzpfO^2\gammaiO/4)$.  In these units the mechanical mode occupancy $\left(\nbar\right)$ and the phonon imprecision level ($\nimp$) can be simply read off from the peak height of the $940$~kHz resonance and the nearby background level, respectively~\cite{cohen_optical_2013}. A plot of the measured $\nimp$ versus $\ncavO$ is shown in \Figref{Fig_SQL}b as grey circles.  The expected imprecision due to vacuum noise of the signal beam is given by $\nimpSN = \kappa\gammaiO/(64\ncavO\gzeroO^2\etaT)$, which is plotted in Fig.~\ref{Fig_SQL}b with no free parameters for the measured $\etaT=0.27$ (solid cyan curve).  Also plotted are the theoretical quantum back-action due to the shot noise of the signal beam ($\nBASN$; solid red curve) and the quantum-limited total added measurement noise (solid green curve). The minimum total added measurement noise assuming $\nBA=\nBASN$ occurs at a signal power corresponding to $\ncavO=0.12$, and represents the standard quantum limit (SQL) for our measurement set-up, $\nsql = 1/(2\sqrt{\etaT})=0.96$ quanta~\cite{cohen_optical_2013}.  The imprecision is vacuum noise limited for all but the highest powers ($\ncavO \gtrsim 500$), reaching a value $34$~dB below the SQL imprecision for an ideal detector (=$1/4$ quanta~\cite{Teufel2009}).  For all the measurement powers shown, the back-action noise is bounded ($\nBA < 4\times10^5$) by the standard deviation $(\pm 3\%)$ in our measurement of the large thermal noise in the mechanical resonator.

\begin{figure}[t]
\includegraphics[width=\columnwidth]{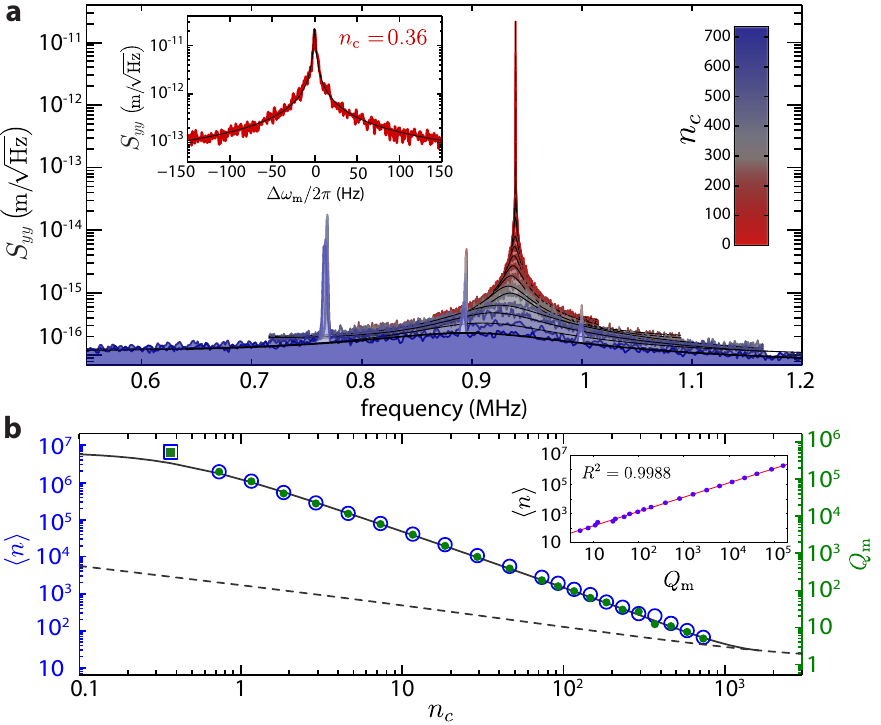}
\caption{\textbf{Room-temperature laser cooling.} \textbf{a}, Plots of the transduced mechanical spectra with the optical feedback engaged for increasing $\ncavO$, showing the damping and cooling of the dominant mechanical mode. Color scale indicates photon number of each mechanical spectrum. The black curves are the fits to the measured spectra used to extract the damped mechanical $Q$-factor ($\QmO$) and phonon occupation ($\nbar$). Spectral peaks at $710$~kHz and $910$~kHz are weakly transduced mechanical modes, while that at $1$~MHz is extraneous electronic noise. Inset: room-temperature mechanical spectra taken with the feedback off for calibration. \textbf{b}, Plot of the inferred $\nbar$ (blue circles; left axis) and $\QmO$ (green solid circles; right axis) extracted from the fits to spectra shown in (a). The first data point (square), at lowest $\ncavO$, is taken from the data in the inset of (a) with the feedback off and is used as a room-temperature calibration. The $y$-axes scales are normalized such that ideal cold-damping would result in the blue and green data points lying directly on top of each other. Solid black curve is the theoretical prediction of cooling with no fit parameters and the dashed line is the minimum possible cooling for our system if the circuit gain is optimized at each $\ncavO$(see App.~\ref{GFBCsection}). The uncertainty in the inferred $\nbar$ of $\pm 15\%$ is smaller than the data points and is dominated by the $95\%$ confidence interval of the fits to each mechanical spectra.  Inset shows the measured phonon occupation versus mechanical quality factor during the cooling run (blue points).  A linear fit to the data is shown as a red curve with R squared value of $0.9988$.} \label{FBC_RT}
\end{figure}

Feedback cooling of the main $940$~kHz mechanical resonance is performed as follows.  As in the open loop measurements, the laser is tuned on resonance with the optical cavity ($\Delta = 0$) and the LO phase is locked to $\thetaH = \pi/2$ which maximizes the measured BHD signal due to mechanical motion.  The BHD signal ($\xmeas (t)$) is fed to an electronic feedback circuit which modulates the intensity of the probe laser incident on the cavity approximately in proportion to $-\xmeasdot$.  The mechanical resonator responds to the intensity modulations of the incident probe laser, imprinting its motion in the orthogonal phase quadrature of the reflected probe light for $\Delta = 0$~\cite{safavi-naeini_squeezed_2013}.  As desired then, for this configuration ($\Delta = 0$, $\thetaH = \pi/2$) it is only the response of the mechanical resonator to the feedback which is recorded in the BHD signal.  The electrical gain of the feedback circuit is held fixed at a value found to yield maximum cooling at the highest value of $\ncavO$ (=$734$), and the laser probe power is increased from low to high, increasing the total loop gain, and thereby increasing the observed cooling and damping.  

Figure~\ref{FBC_RT}a shows the measured NPSD around the main $940$~kHz mechanical resonance with optical feedback applied as per the above prescription.  At each measured optical power, the mechanical spectra are fit with a Lorentzian curve (black solid lines) from which an area and linewidth of the spectrum are determined.  The phonon occupancy of the mechanical resonance, plotted as blue circles in \Figref{FBC_RT}b, is proportional to the transduced area under the mechanical spectrum normalized by $\ncavO$, whereas the damped mechanical $Q$-factor is determined from the linewidth. Absolute calibration of the phonon occupancy for each of the optical powers is found by comparing to the measured area under the mechanical spectrum at the lowest power point ($\ncavO =0.36$) with the feedback off.  At this power, and in open loop, dynamic back-action effects are negligible and the mechanical resonance is at the room temperature thermal occupancy ($\Tbath=295$~K, $\nbar = \nbath = 6.5 \times 10^6$). Of crucial importance to the interpretation of the data presented in \Figref{FBC_RT}, is the fact that the change in the mechanical quality factor (green dots) follows the change in the measured occupation (see inset), which as discussed earlier, is a hallmark of feedback cooling.

The lowest phonon occupation achieved here is $\nbar = 66 \pm 10$.  At this cooling point the linewidth of the mechanical resonance at $\omegamO/2\pi = 940$~kHz has been broadened to $190$~kHz.  Further cooling of the mechanical resonance is limited primarily by the combination of two sources. An excess time delay of $1 \mu$s in our feedback loop modifies the broadband phase response of the system away from the ideal value of $\thetafb = 0$, leading to amplification, rather than damping, of mechanical motion for frequencies outside an approximate $250$~kHz bandwidth around the mechanical resonance (see App.~\ref{sec:exp_cal}).  Additionally, at the highest optical cooling power of $\ncavO = 734$ we are nearing an optical-absorption-induced, thermo-optic bistability of the optical cavity response.  Calibration of the absorption heating via the thermo-optic tuning of the optical cavity resonance indicates only a $\Delta T \approx 10$~K increase of the local bath temperature for $\ncavO = 734$, representing a parasitic back-action noise of $\nBAabs \approx 300\ncavO$. Although $\nBAabs \ll \nbath$ at this optical power, which doesn't impact the current level of cooling, as one approaches mode occupancies of $\nbar \lesssim 1$ the back-action noise becomes relevant.      

Cooling to the quantum ground-state from room temperature remains an achievable goal, but requires improvement in several key device properties.  As discussed in Ref.~\cite{genes_ground-state_2008}, one cannot increase the feedback bandwidth without limit as eventually the amount of imprecision noise (white shot noise in the ideal case) fed back onto the mechanical resonator is enough to heat it out of the ground-state.  As shown in App.~\ref{GFBCsection}, consideration of the feedback bandwidth in turn constrains the intrinsic mechanical $Q$-factor to a value greater than the thermal bath occupancy ($\QmO \gtrsim 3\nbath$).  Quality factors approaching this limit for MHz-frequency resonators in thin film $\SiN$ have been achieved by modifying the structure to minimize losses at the clamp points~\cite{schmid_damping_2011,yu_control_2012}, utilizing nitride with fewer bulk defects and higher stress~\cite{faust_signatures_2013}, and modifying the post-etching surface properties of the nitride~\cite{KS_private}.  Even with an increase in the mechanical $Q$-factor, however, reaching the ground-state with the optomechanical coupling strength of the devices in this work would still require a prohibitively large intra-cavity photon number of $\ncavO \approx 5\times10^{4}$. By increasing the optomechanical coupling and optical $Q$-factor to levels previously demonstrated in similar devices ($\gOMO/2\pi = 200$~GHz/nm, $\QoO = 6\times 10^4$~\cite{camacho_characterization_2009}), this can be lowered to a value as small as $\ncavO \approx 60$.  There is also the issue of the parasitic back-action.  In the current devices we estimate an imprecision-back-action product of $\nimp \nBA \approx 450 (\nimp \nBA)_{\text{H}}$ at $\ncavO = 734$, where $(\nimp \nBA)_{\text{H}}=1/16$ is the Heisenberg quantum limit.  With the above mentioned device improvements, $\nBASN \gg \nBAabs$, and the imprecision-back-action product would approach $1/\etaT \approx 3.7$ of the Heisenberg limit, suitable for cooling below unit occupancy. 

Viewed from the perspective of a continuous position measurement, attaining ground-state cooling requires (approximately) that the rate at which measurement information is gained about the mechanical motion ($\Gammameas = 4\eta_t\ncavO\gzeroO^2/\kappa$) be greater than the rate at which the mechanical resonator is disturbed by coupling to its thermal bath ($\Gammath = \gammaiO(\nbath+1)$), and approach the back-action decoherence rate ($\Gammadeph = \gammaiO \nBA$)~\cite{Clerk_RMP_2010,miao_achieving_2010,Wilson2014}.  More generally then, the photonic crystal optomechanical devices studied here could enable quantum measurement and control protocols~\cite{Wiseman1994,wiseman_using_1995,Doherty1999,clerk_back-action_2008,vanner_selective_2011,chen_macroscopic_2013,szorkovszky_strong_2013} for the preparation of mechanical objects in highly non-classical quantum states of motion.  With the device improvements mentioned above, these protocols could be implemented without additional cryogenic cooling, and in a room temperature environment where they may be employed for a variety of precision sensing applications.  In a similar vein, feedback control is commonly employed in MEMS sensors of forces and fields~\cite{gorman_feedback_2011,Miao2012} to change the frequency, bandwidth, and dynamic range of the sensor.  In this work the bandwidth of the resonator's response is increased from $1.7$~Hz to $190$~kHz and the dynamic range is increased by $50$~dB, all while preserving the undamped thermal noise force sensitivity of $125$~aN/Hz$^{1/2}$.  When applied to the field of atomic force microscopy, for example, such an optomechanical sensor~\cite{Srinivasan2011} might be used to improve imaging resolution by reducing thermal motion of the sensor tip~\cite{garbini_optimal_1996}, or in the case of measurements of molecular motion, to resolve dynamics at microsecond time scales~\cite{ando_high-speed_2008}. 

\begin{acknowledgments}
The authors would like to thank Dal Wilson and Kartik Srinivasan for helpful discussions. This work was supported by the DARPA QuASaR program through a grant from the Army Research Office, the Institute for Quantum Information and Matter, an NSF Physics Frontiers Center with support of the Gordon and Betty Moore Foundation, and the Kavli Nanoscience Institute at Caltech. TDB gratefully acknowledges support from the National Science Foundation Graduate Research Fellowship Program (grant no. 0703267). 
\end{acknowledgments}


\appendix

\section{Device characterization}
\label{sec:dev_char}

\subsection{Optical mode characterization}
\label{subsec:opt_char}

\begin{figure}[t]
\includegraphics[width=\columnwidth]{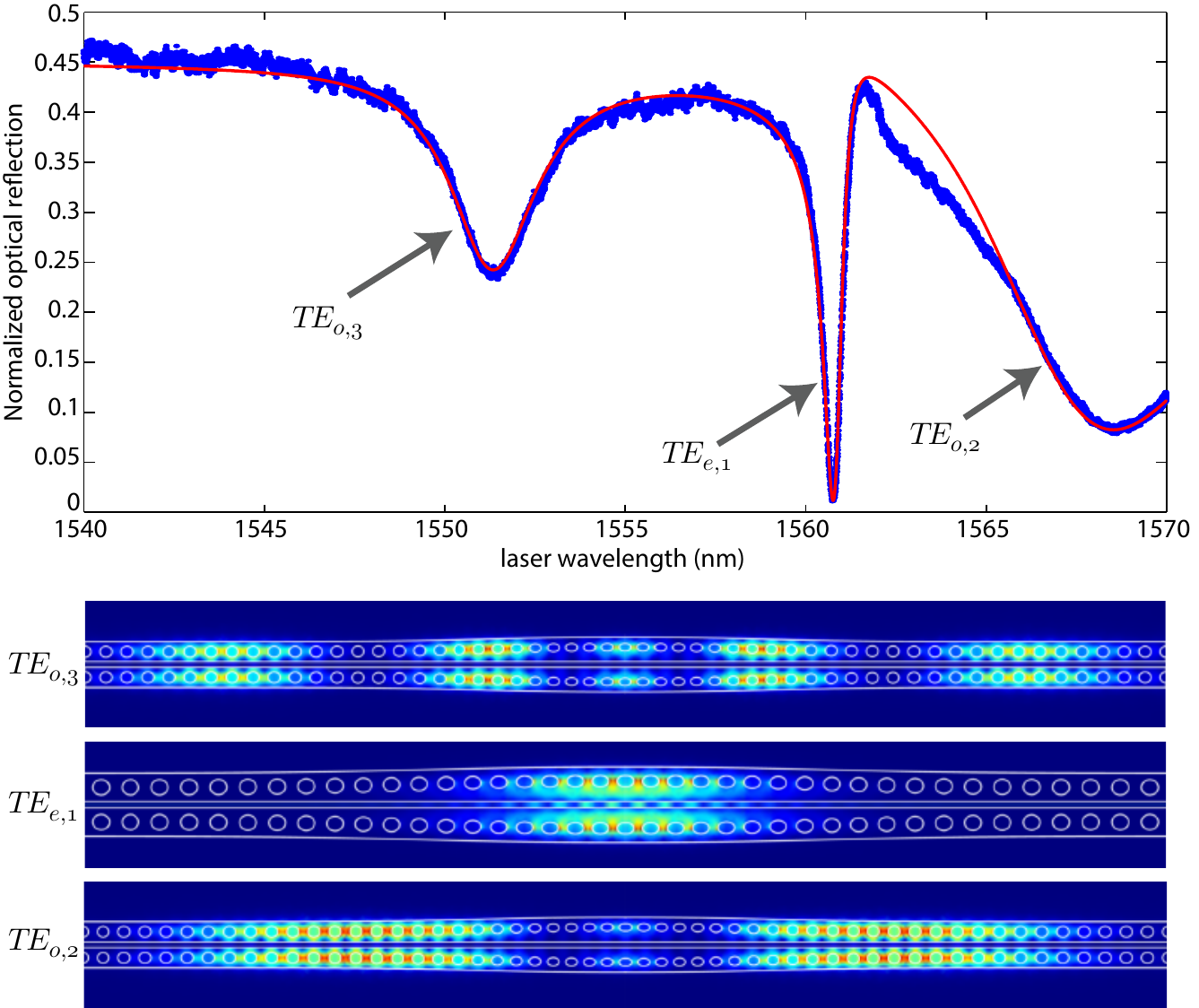}
\caption[]{Normalized optical reflection spectrum of the device studied in the main text, measured by scanning the probe laser wavelength from $\lambda=1540$-$1570$~nm.  The signal level is the ratio of the measured reflected optical laser power to the optical power incident at the optical cavity. The blue curve is the measured laser wavelength scan and the red curve is a fit used to determine the optical coupling efficiency between the collection fiber and the on-chip waveguide, as well as the optical decay rate ($\kappa$) and the waveguide-to-cavity coupling rate ($\kappae$) of each optical cavity mode. The images below the normalized optical reflection scan are the numerical simulations of the energy density of the identified optical modes.  The simulation images are labeled as $\Omodej$, where the subscript $e(o)$ indicates the spatial mode symmetry about the $\sigma_{x}$ mirror plane and the subscript indicates the order of the mode along the waveguide axis.}\label{SI_trans}
\end{figure}

The photonic crystal cavity used in this work exhibits a series of optical resonances of different symmetries and orders. A detailed description of a photonic crystal with a similar structure can be found in Ref.~\cite{camacho_characterization_2009}. The main symmetry of interest ($\sigma_{x}$) is about a plane which runs parallel to the length of the beams and is situated in the center of the gap between them. The optical field spatial mode pattern can be either even ($e$) or odd ($o$) about this symmetry plane, where $e$ modes have their fields concentrated in the gap and the field of $o$ modes experience a node at the center of the gap. Due to their higher field concentrations in the gap, the resonance frequencies of $e$ modes are more sensitive to the in-plane differential motion of the two beams, $x$, than are the $o$ modes.  As a result the $e$ modes tend to have larger optomechanical coupling to the in-plane differential motion of the nanobeams. By measuring how well an optical mode transduces differential motion of the beams we can determine its symmetry. Accordingly, the resonances at $\lambda=1551$~nm and $1568$~nm are labeled as odd symmetry, whereas the optical mode of interest at $\lambda=1561$~nm is of even symmetry.  In addition the optical modes are labeled by an integer subscript $j=0,1,2,...$ related to the number of nodes in the spatial mode pattern of the optical mode along length of the beams.  The order of the optical modes are determined by comparing the measured optical frequency to that found from COMSOL Multiphysics FEM simulations~\cite{COMSOL}.

Figure~\ref{SI_trans} shows the laser reflection spectrum of the device ``zipper'' cavity device studied in the main text.  There are three optical resonances measured in the $1550$~nm telecom wavelength band, which we identify and label as in the simulated plots shown below the measured reflection spectrum in Fig.~\ref{SI_trans}.  The resonance at $\lambda=1561$~nm is the one used for feedback cooling in the main text. Considering the Fano-like lineshape of the resonances, we fit the wide optical scan (blue curve) to a model consisting of three overlapping optical resonances and find good agreement (red curve). The fit to the $\Omode$ mode of interest at $1561$~nm yields $\omegacO/2\pi = 192$~THz, $\QoO = 2500$, $\kappaO/2\pi = 77$~GHz, and $\kappae/\kappa=0.63$, where $\omegacO$ is the angular frequency of the optical cavity, $\QoO$ is the quality factor, $\kappaO$ is the total angular loss rate, and $\kappae$ is the angular coupling rate to the waveguide.

\subsection{Calibration of the optomechanical coupling strength}
\label{subsec:cal_gOM}

\begin{figure}[t]
\includegraphics[width=0.7\columnwidth]{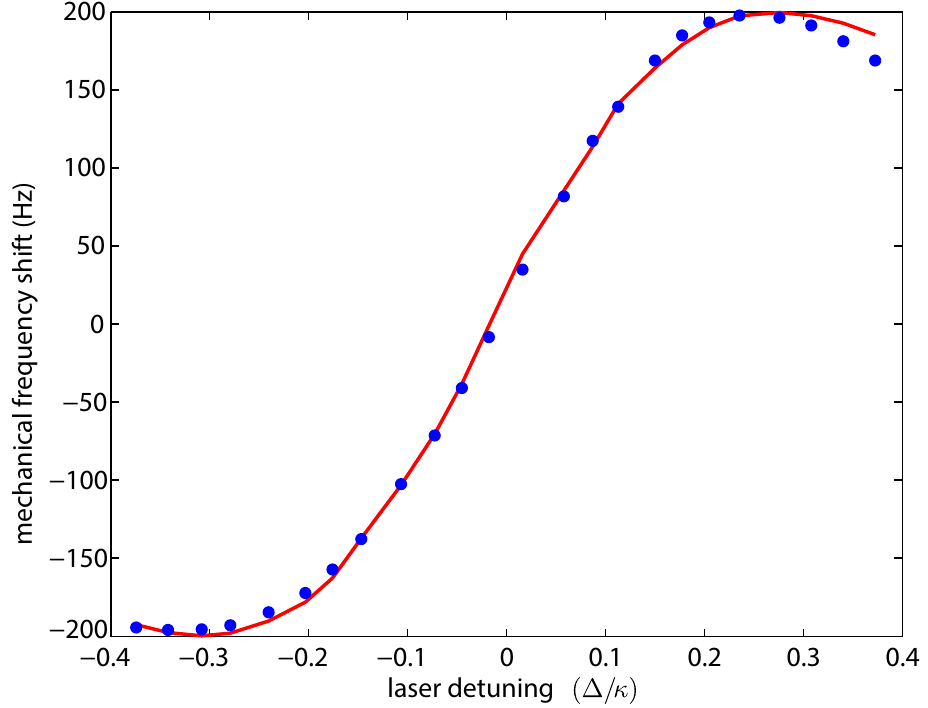}
\caption[]{Measurement of the mechanical spring shift for a range of laser-cavity detunings (blue circles).  The theoretical frequency shift for an optomechanical coupling coefficient of $\gOMO/2\pi= 36$~GHz/nm ($\gzeroO/2\pi = 358$~kHz) is shown as a red curve.}\label{SI_SpringGom}
\end{figure}

In order to calibrate the optomechanical coupling coefficient, $\gOMO \equiv d\omegacO/dx$, we measure the optically-induced spring shift of the $\Mmode$ mechanical resonance versus laser-cavity detuning $\Delta$.  In the sideband unresolved regime in which $\omegam \ll \kappa$ the mechanical frequency shift is given by~\cite{eichenfield_picogram-_2009},

\begin{align}
\omegamO & = \omegamiO \left[1 + \left(\frac{2\hbar \gOMO^2 \ncavO}{\meff\omegamO^2}\right)\frac{\Delta}{\Delta^2+\left(\frac{\kappa}{2}\right)^2}\right]^{1/2} \nonumber \\
	& \approx \omegamiO + \left(\frac{\hbar \gOMO^2 \ncavO}{\meff\omegamO}\right)\frac{\Delta}{\Delta^2+\left(\frac{\kappa}{2}\right)^2}.
\end{align}

\noindent where $\omegamiO$ is the bare mechanical frequency, $\ncavO$ is the number of intracavity photons, $\meff$ is the effective motional mass, $\kappa$ is the optical cavity (energy) decay rate. In Fig.~\ref{SI_SpringGom} we show the measurement of the spring shift (blue data) and the fit curve (red line), from which we find the vacuum coupling rate to be $\gzeroO \equiv \gOMO\xzpfO = 2\pi[358$~kHz$]$.  The zero-point fluctuation amplitude $\xzpfO=\left(\hbar/2\meff\omegam \right)^{1/2} = 9.7$~fm of the $\Mmode$ mechanical mode is determined from the mechanical frequency and the simulated motional mass, $\meff = 90 \times 10^{-15}$~kg~\cite{eichenfield_modeling_2009}.  This yields a linear dispersive coupling coefficient of $\gOMO/2\pi = 36$~GHz/nm from the vacuum coupling rate, in good agreement with the simulated value from COMSOL \cite{COMSOL} of $\gOMO/2\pi = 41$~GHz/nm.

\subsection{Measurement of mechanical quality factor}
\label{subsec:Qm}

\begin{figure}[t]
\includegraphics[width=0.7\columnwidth]{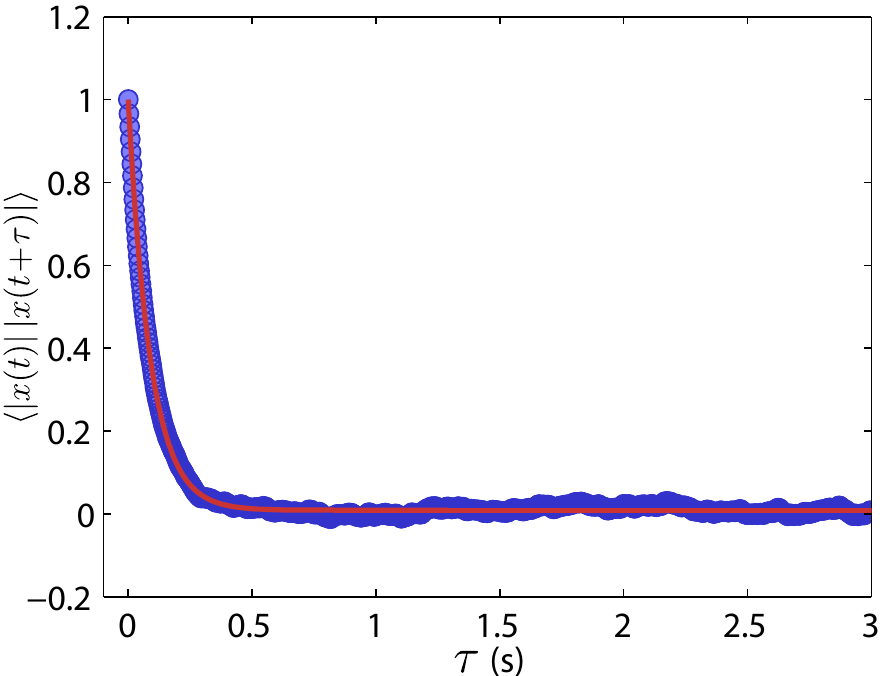}
\caption[]{Measurement of the amplitude autocorrelation, $\langle |\tilde{x}(t)||\tilde{x}(t+\tau)| \rangle$, of the thermal motion of the $\Mmode$ mechanical resonance.  The blue data is the measured autocorrelation and the red line is a fit to the exponential decay, yielding a mechanical $Q$-factor of $\QmO=5.3\times10^5$.}\label{SI_ring}
\end{figure}

Here we describe details of the measurement of the intrinsic mechanical quality factor of the fundamental in-plane differential mechanical resonance ($\Mmode$) of the ``zipper'' cavity structure. The approximately $1$~Hz linewidths of this mechanical resonance combined with the minimum resolution bandwidth of our spectrum analyzer (also $1$~Hz) makes measuring the mechanical quality factor from the spectral density difficult. However, the amplitude of a thermally-driven resonator will be correlated with itself over a timescale dictated by the mechanical damping rate corresponding to the coupling of the resonator to the thermal bath. Formally it can be shown that the autocorrelation of the amplitude, $\langle |\tilde{x}(t)||\tilde{x}(t+\tau)| \rangle$, will decay as $\mathrm{e}^{-t/\taum}$, where $\taum = \QmOi/\omegamO$ and $\omegamO$ is the intrinsic mechanical angular frequency~\cite{stipe_noncontact_2001}. The measurements of the thermal amplitude of motion ($x(t)$) are made using the experimental setup described in the main text with the laser on-resonance with the optical cavity and the feedback off. The slowly varying envelope of the mechanical resonator motion, $|\tilde{x}(t)|$, is obtained from the magnitude channel of a lock-in amplifier tuned to the mechanical resonance frequency. We use a bandwidth of $\approx 100$~Hz, much larger than the mechanical linewidth, which ensures that any frequency jitter in the mechanical resonance does not affect the measurement. Figure~\ref{SI_ring} displays the autocorrelation of the measured signal. A exponential curve fit to the decay of the autocorrelation signal yields a coherence time of $\taum=90$~ms, which for $\omegamO/2\pi=940$~kHz gives $\QmOi=5.3\times10^5$. For lower-$Q$ structures it was confirmed that this technique agrees with a direct measurement of the linewidth as measured on a spectrum analyzer. Furthermore, to be sure that we are not doing significant dynamic back-action, and thus that we have obtained the \emph{intrinsic} quality factor, we performed the measurement at the low optical power where there was no power dependence to the measured $\taum$.

\section{Optical feedback cooling: theory} 
\label{GFBCsection}

In this section we present (without derivation) the results from Ref.~\cite{genes_ground-state_2008} for the variances of the position and momentum of a harmonic oscillator in a derivative feedback loop with an on-resonant quantum noise limited laser. First, though, we define the variables used in Ref.~\cite{genes_ground-state_2008}:

\begin{align}
&\gzeroO = \gOMO \xzpfO, \\
&G = \gzeroO \sqrt{\ncavO}, \\
&\nimpSN \ \frac{\kappaO\gammaiO}{64G^2\etaT}, \\
&\nBASN = \frac{4G^2}{\kappaO\gammaiO}, \\
&g = \frac{4 \geecd G \omegamO}{\kappaO\gammaiO}, \label{genesvariables}
\end{align}

\noindent where $G$ is the parametrically-enhanced coupling rate between the optics and the mechanics, $\ncavO$ is the average intracavity photon number of the probe laser, $\nimpSN$ is the shot-noise-limited imprecision in units of phonon number, $\nBASN$ is the quantum back-action of the shot noise in units of phonon number, $\etaT$ is the total quantum detection efficiency~\cite{cohen_optical_2013}, $g$ is a normalized unitless feedback strength, equivalent to that used in the main text, and $\geecd$ is a unitless gain term accounting for the feedback circuit response.

The feedback response is taken to be a standard derivative high-pass filter with cut-off frequency $\omegafb$, which in the Fourier domain is given by $\mathcal{F}(\omega) = -i\omega\geecd/\left(1-i\omega/\omegafb\right)$.  With this assumed feedback response function the variances of the two quadratures of the mechanical mode are given by (under certain assumptions~\cite{genes_ground-state_2008} valid in this work),

\begin{widetext}
\begin{align}
\langle \delta q^2 \rangle = \left[g^2\nimp +\left(\nbath +\frac{1}{2}+\nBA \right)\left(1+ \frac{\omegamO^2}{\omegafb^2} \right)\right] \left(1 + g + \frac{\omegamO^2}{\omegafb^2}\right)^{-1}, \label{q2}
\end{align}

\begin{align}
\langle \delta p^2 \rangle =  \left[g^2\nimp\left(1  + \frac{g\gammaiO\omegafb}{\omegamO^2} \right) + \left( \nbath + \frac{1}{2} + \nBA \right)\left( 1 + \frac{\omegamO^2}{\omegafb^2} + \frac{g\gammaiO}{\omegafb} \right)  \right] \left(1 + g + \frac{\omegamO^2}{\omegafb^2}\right)^{-1}. \label{p2}
\end{align}
\end{widetext}

\noindent To determine the fundamental cooling limits from these equations we take the limit of large feedback bandwidth $\left(\omegamO/\omegafb\right)^2 \ll 1$ and large feedback strength $g \gg 1$, which allows us to drop some terms, and write a simpler formula for the position fluctuations,

\begin{equation}
\langle \delta q^2 \rangle =  g\nimp + \frac{\nBA}{g} + \frac{\nbathhalf}{g}. \label{qsimple}
\end{equation}

\noindent Taking the same limits for the momentum variance we find,

\begin{multline}
\langle \delta p^2 \rangle = \left[g\nimp + \frac{\nBA}{g}  +  \frac{\nbathhalf}{g}\right] \\ 
+ \gammaiO\left[ \frac{g^2\nimp\omegafb}{\omegamO^2} + \frac{1}{\omegafb}\left(\nbath + \nBA \right) \right]. \label{pvar}
\end{multline}

\noindent Here we note that the first bracketed term on the RHS of the formula for the momentum variance looks the same as that for the position variance in \Eqref{qsimple}. For now we will assume that the second bracketed term on the RHS is small and ignore it. We will revisit this assumption at a later point and determine when this assumption is valid. To be explicit, our working assumption is stated below:

\begin{align}
\gammaiO\left[ \frac{g^2\nimp\omegafb}{\omegamO^2} + \frac{1}{\omegafb}\left(\nbath + \nBA \right) \right] < 1. \label{pRHSminimize}
\end{align}

\noindent This leaves us with the simplified (and approximate) form of the momentum variance under optical feedback cooling, 

\begin{align}
\langle \delta p^2 \rangle & =  g\nimp + \frac{\nBA}{g} + \frac{\nbathhalf}{g} \label{psimple}.
\end{align}

In order to relate these variances to a phonon occupation number, we equate the total energy of the oscillator to the sum of its variances, 

\begin{equation}
E = \frac{\hbar \omegamO}{2}\left[ \langle \delta q^2 \rangle + \langle \delta p^2 \rangle \right] = \hbar \omegamO \left( \nbar + \frac{1}{2}\right) \label{qp2n}
\end{equation}

\noindent which yields for the average mode occupation,

\begin{equation}
\label{nbar1}
\nbar + \frac{1}{2}  = \frac{1}{2}\left[ \langle \delta q^2 \rangle+ \langle \delta p^2 \rangle \right].
\end{equation}

\noindent Substituting the simplified formulas for the variances (\Eqref{qsimple} and \Eqref{psimple}) gives eqn.~(2) of the main text, 

\begin{equation}
\nbar + \frac{1}{2} \approx g\nimp + \frac{\nBA}{g} + \frac{\nbathhalf}{g}. \label{eq2maintext}
\end{equation}

\begin{figure*}[htb]
\includegraphics[width=1.5\columnwidth]{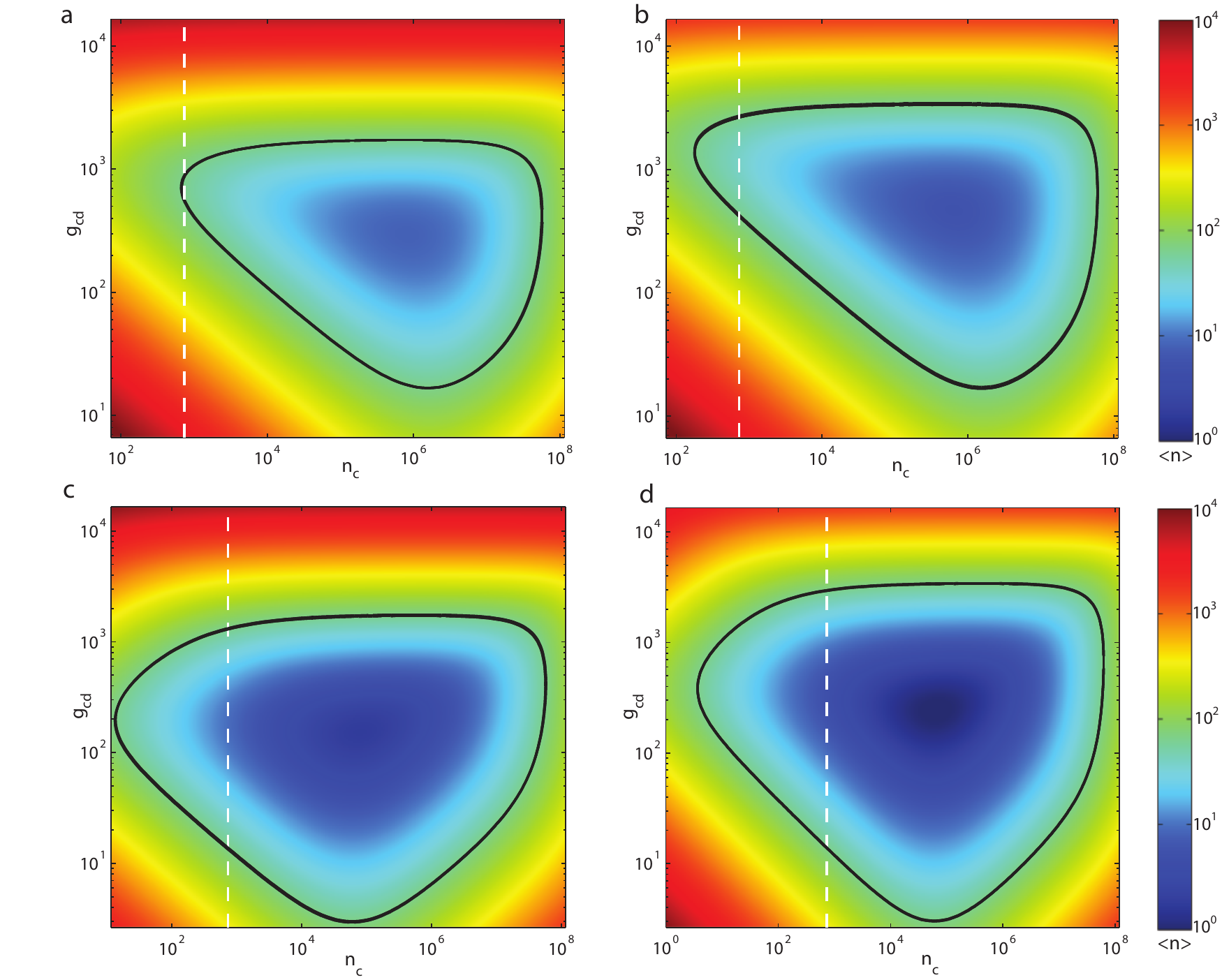}
\caption[]{\textbf{a}, Color density plot of $\nbar$ for ideal radiation-pressure derivative feedback cooling using the device parameters of the main text, versus normalized gain and photon number. The black line is a contour at $\nbar = 66$, the minimum phonon number achieved in the main text. The white dashed line denotes $\ncavO=734$, which is the maximum number of photons that can stably occupy the cavity. The minimum occupation is $\nbarmin = 64$ for $\ncavO=734$. The global minimum at the ideal power and gain is $\nbarmin = 8$. \textbf{b}, Same as (a) except with perfect detection efficiency $\etaT=1$. The minimum occupation is $\nbarmin = 38$ for $\ncavO=734$. The global minimum at the ideal power and gain is $\nbarmin = 6$. \textbf{c}, Same as (a) except with $\QmOi = 1.5\times 10^7$. $\nbarmin = 9.4$ for $\ncavO=734$ and $\nbarmin = 1.8$ for the ideal power and gain. \textbf{d}, Same as (c) except with $\etaT=1$. $\nbarmin \cong 5$ for $\ncavO=734$ and $\nbarmin =0.95$ for the ideal power.}\label{FB_theory_Master}
\end{figure*}

The optimal $g$ that minimizes $\nbar$ in eqn.~(\ref{eq2maintext}) is $\gopt = \left[\left(\nbath + \nBA\right)\nimp\right]^{1/2}$, which upon re-substitution yields a minimum phonon occupancy of, 

\begin{align}
\label{nbarmin}
\nbarmin + \frac{1}{2}  & = 2\sqrt{\nimp\left( \nbath + \nBA \right)} \\
                        & \approx 2\sqrt{\nimp\nbath + \frac{1}{16\etaT}},
\end{align}

\noindent where in the final expression we have used the quantum-limited relation, $\nBASN = 1/(16\etaT\nimpSN)$. From this final result we can establish the two requirements mentioned in main text for reaching $\nbar < 1$, 

\begin{align}
 \etaT & > \frac{1}{9},\\
 \nimp & < \frac{1}{16\nbath}\left(9 - \frac{1}{\etaT}\right). \label{nimp_needed}
\end{align}

\noindent Of course other technical limitations may come into play, such as excess back-action or instabilities which limit the optical power and thus feedback gain.  Furthermore, we must be in a parameter regime that satisfies the assumption we made at the outset in \Eqref{pRHSminimize}. Substituting $\gopt$ into \Eqref{pRHSminimize} yields the following relation, 

\begin{equation}
\label{Qmlimit}
\QmOi > \left(\nbath+\nBA\right)\left(\frac{\omegafb}{\omegamO}\right),
\end{equation}

\noindent where we have again used $\left(\omegamO/\omegafb\right)^2 \ll 1$. This relation can be further simplified by saturating the inequality in \Eqref{nimp_needed} and relating $\nimp$ to $\nBA$ through their quantum-limited relation, 

\begin{equation}
\label{Qmlimit_simple}
\QmOi > \left(\nbath+\frac{\nbath}{9\etaT-1}\right)\left(\frac{\omegafb}{\omegamO}\right) .
\end{equation}

\noindent If we further assume ideal detection ($\etaT = 1$) and take the feedback bandwidth to be $\omegafbopt \approx 3 \omegamO$, a value numerically found to be the optimal feedback bandwidth in most cases~\cite{genes_ground-state_2008}, this yields the requirement stated in the main text: $\QmOi \gtrsim 3\nbath$ in order to reach a cooled phonon occupation of $\nbar \lesssim 1$ using optical measurement and radiation pressure feedback.

We can also use the full formulas, \Eqref{q2} and \Eqref{p2}, to predict the maximum cooling for a given set of parameters. Figure~\ref{FB_theory_Master}a shows the theoretical cooling plot for the parameters of the device studied in this work. Figure~\ref{FB_theory_Master}b shows the same cooling plot but for unit detection efficiency ($\etaT=1$).  From these plots we see that for a maximum intracavity photon number, $\ncavO = 734$, the best achievable for the device parameters in work and assuming ideal derivative feedback cooling is $\nbarmin = 64$, very close to the measured minimum mode occupancy of $\nbar = 66 \pm 10$.  With the limited detection efficiency of $\etaT=027$, the minimum achievable occupation is $\nbarmin = 8$.  For perfect detection efficiency $\nbarmin = 6$ could be reached. Both of these estimates have neglected excess back-action.  Furthermore, achieving this level of cooling requires using very high intracavity photon numbers, which would cause thermal instabilities in our system.  As discussed in the main text, several device improvements would be required to achieve this level of cooling.  Figures.~\ref{FB_theory_Master}c and d show cooling plots assuming an improved mechanical $Q$-factor of $\QmOi = 1.5 \times 10^7$.  

\section{Detection and feedback calibration}
\label{sec:exp_cal}

\subsection{Determination and locking of homodyne phase}
\label{subsec:PhaseLockSection}

\begin{figure}[htb]
\includegraphics[width=\columnwidth]{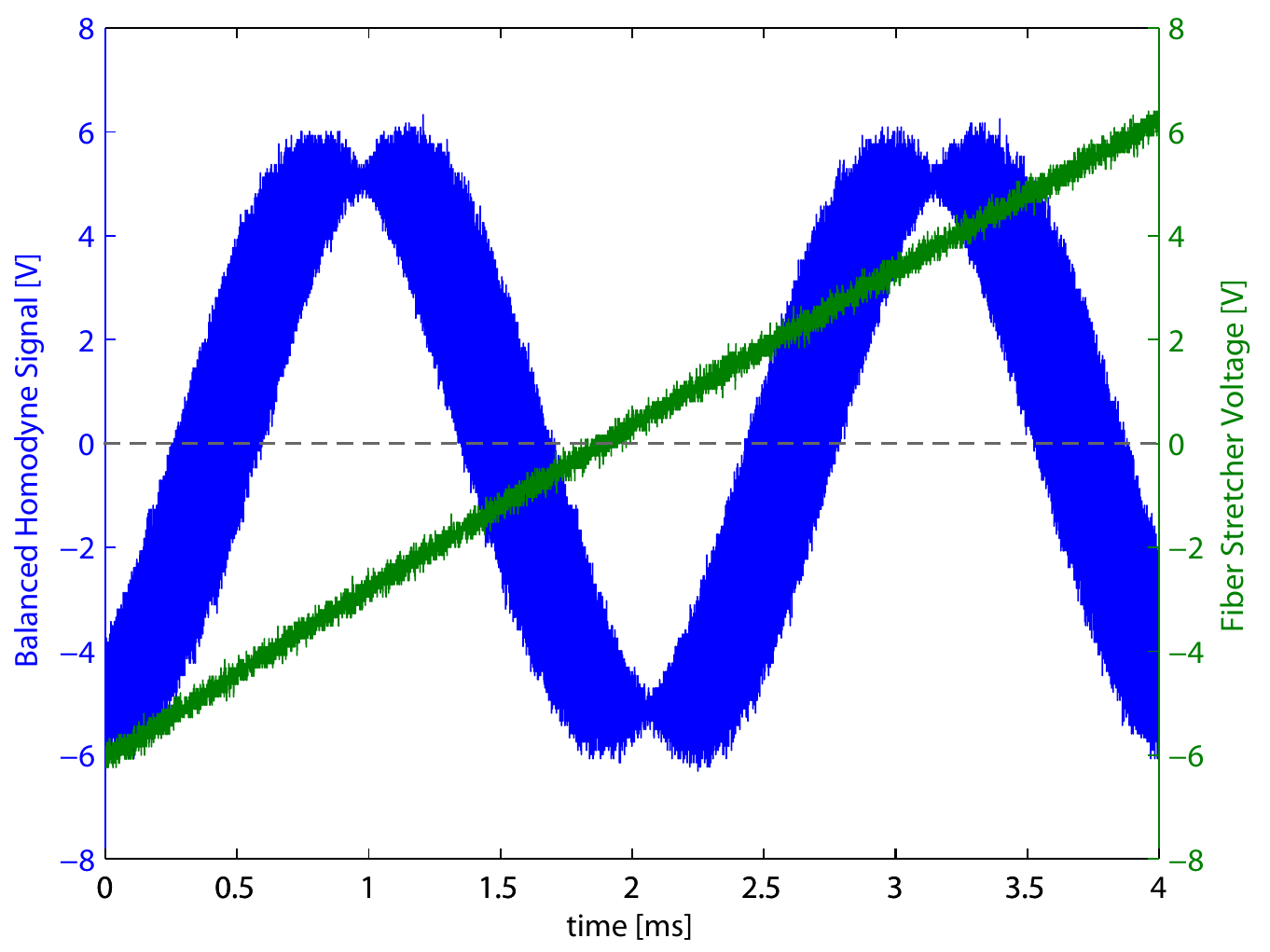}
\caption[]{Plot of a balanced homodyne signal (blue) as the fiber stretcher voltage is swept (green), with the laser on-resonance with the optical cavity.}\label{SI_homodyne_trace}
\end{figure}

\begin{figure*}[htb]
\includegraphics[width=1.7\columnwidth]{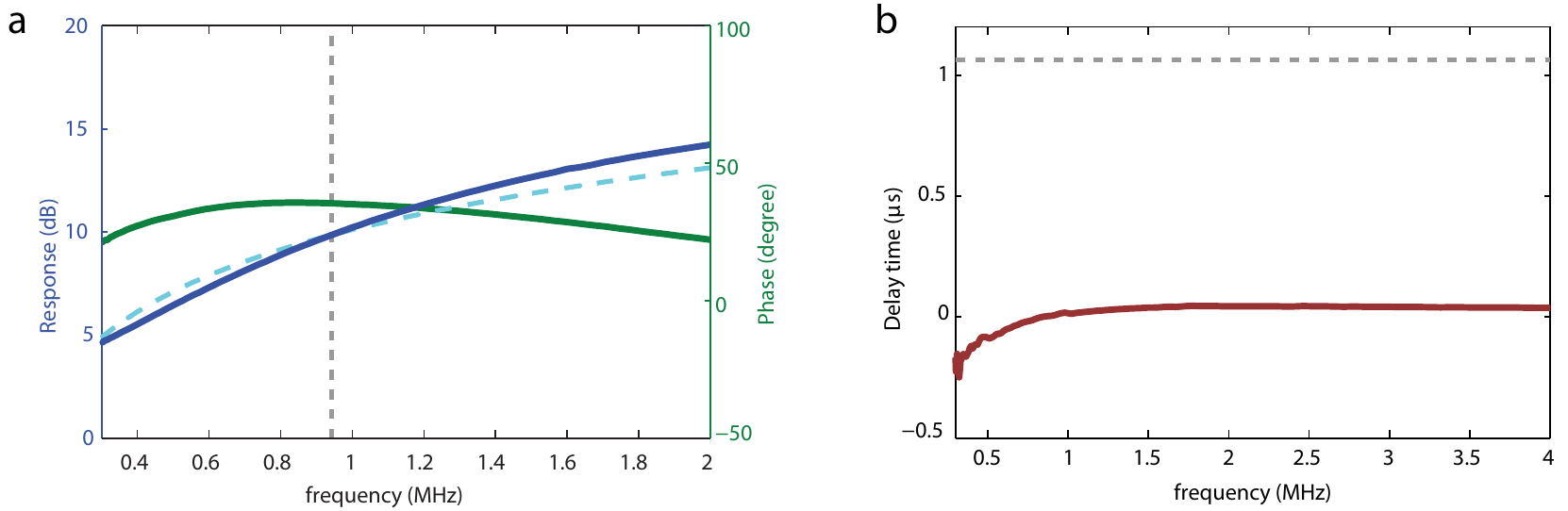}
\caption[]{\textbf{a}, Magnitude (blue) and phase (green) response of the mFALC110 for the settings used in the measurements presented in the main text. Vertical grey dashed line indicates the mechanical frequency and the cyan dashed line indicates the amplitude response for ideal derivative feedback, proportional to $\omega$. \textbf{b}, Total time delay (red) as measured on a network analyzer. Horizontal grey dashed line indicates one mechanical period.}\label{SI_WideOffRes1}
\end{figure*}

The relative phase angle between the local oscillator (LO) and the reflected cavity signal ($\thetaH$) determines which optical quadrature of the reflected signal one measures at the output of the BHD.  One can easily show that for an optical probe on-resonance with the optical cavity ($\Delta=\omegaL-\omegacO=0$) the position of the mechanical resonator is imprinted on the phase quadrature of the reflected light corresponding to $\thetaH = \pi/2$~\cite{safavi-naeini_squeezed_2013}.  This phase angle is set by the relative path lengths of the LO and signal arms, and slow thermal drifts or acoustic noise can cause changes in the relative path length, thus changing the phase angle and the measured optical quadrature. Note that to avoid complications of laser phase noise, the LO and signal beams are derived from the same laser and the two optical paths are roughly matched in length.  In order to stabilize $\thetaH$ the low-frequency component of the BHD signal ($f<200$~kHz) is fed to a digital feedback system (Toptica DigiLock) that has a high-voltage output panel ($V_{\text{out}} \leq 150$~V) which drives a voltage-controlled fiber stretcher (Optiphase PZ1). This fiber stretcher (FS) consists of a long distance of optical fiber ($\sim 10$~m) wrapped around a bulk piezo element.

Shown in \Figref{SI_homodyne_trace} is the homodyne output while linearly driving the FS and for optical probing at $\Delta=0$. This voltage sweep (green) modulates the detected phase quadrature of the reflected light, which we detect as a DC signal with sinusoidally modulated voltage output proportional to $\cos{\thetaH}$. The thermal Brownian motion of the strongly coupled mechanical mode at $\fmO = 940$~kHz is evident in the blue curve as increased noise on the signal, which decreases rapidly at the maximum and minimum points of the curve corresponding to the read out of the intensity quadrature of the reflected optical signal.  Read-out of the phase quadrature of the reflected light ($\thetaH = \pi/2$) occurs at the average voltage between the peak and trough of this curve (grey dashed line), which is the lock-point desired in our measurements.  In the cooling curve shown in the main text, for each optical probe power the laser-cavity detuning and the homodyne lock-point are adjusted before each each mechanical spectrum is measured, and then checked after the measurement is completed and before moving to the next optical power point. 
\subsection{System response and delay times}
\label{subsec:system_response}
Here we report the system response and delay times of our feedback circuit used to apply radiation pressure feedback. In \Figref{SI_WideOffRes1} we present the measured amplitude, phase response, and time delay of the analog circuit, mFALC110, used as part of the feedback loop to perform the derivative feedback. The cyan dashed line in Fig.~\ref{SI_WideOffRes1}a indicates the amplitude response for purely derivative feedback, which is proportional to $\omega$. Over the frequency range used for this experiment ($0.8$ - $1.2$~MHz), the amplitude response deviates from the ideal slope by $\approx 0.5$~dB and the phase changes by $\approx 5^\text{o}$. The time delay from the mFALC110 circuit, shown in \Figref{SI_WideOffRes1}b, is negligible when compared to the mechanical period.

\begin{figure*}[htb]
\includegraphics[width=1.7\columnwidth]{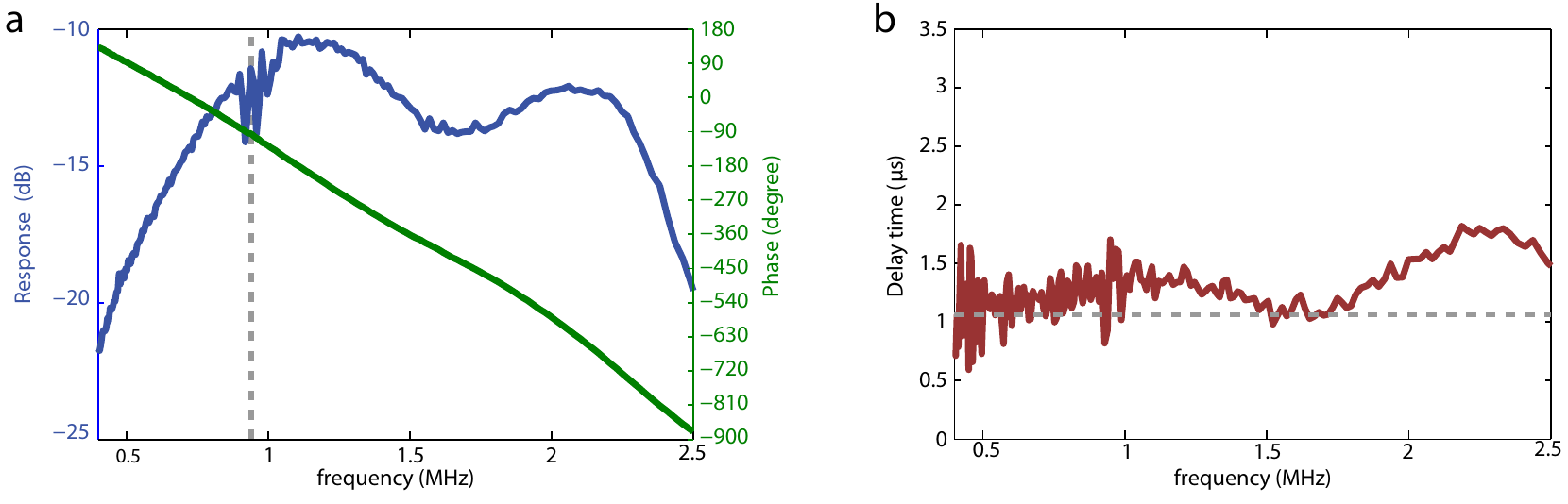}
\caption[]{\textbf{a}, Magnitude (blue) and phase (green) response of the entire feedback system including electronics and optical train, with the laser off resonance from the optical cavity. The grey dashed line indicates the mechanical frequency.  The noisy signal centered at the mechanical frequency is simply a result of residual mechanical transduction and is not indicative of a rapidly changing response of the circuit. \textbf{b}, Total time delay as measured on a network analyzer (red). A single mechanical period is indicated by the horizontal grey dashed line.}\label{SI_WideOffRes2}
\end{figure*}

In Fig.~\ref{SI_WideOffRes2} we present the amplitude and phase response of the entire feedback loop, including the optical portion.  It is important to note that this data was taken off-resonance from the optical cavity ($\Delta/\kappa \gg 1$) so that the mechanical motion would not dominate the response; however, a trace of the mechanical response is still visible in the noise of the curve near the mechanical resonance frequency (vertical grey dashed line).  The BHD phase $\thetaH$ was set to be sensitive to phase fluctuations of the reflected light, as in the experiment. Since the intensity modulator changes amplitude, at this homodyne phase the detected amplitude response is attenuated and should be considered in arbitrary units. Nonetheless, in the frequency range of interest ($0.8$ - $1.2$~MHz) the amplitude response is qualitatively similar to the response of the mFALC110. Of notable difference, however, is that the overall delay time is now comparable to the mechanical period (see horizontal grey dashed line in Fig.~\ref{SI_WideOffRes2}b). This delay time was measured to have the following components: $200$~ns from the optical detector, $400$~ns from the band pass filtering of the feedback circuit, $180$~ns from the phase-shifter, and the remainder is dominated by the total signal path length. The non-negligible delay in the rest of the feedback loop requires the inclusion of the phase shifter which allows us to apply a feedback force that purely damps with little spring shift. A consequence of this nearly constant time delay is that the phase of the feedback signal varies linearly with frequency. This becomes important when considering the wideband feedback loop noise in the next section.

\subsection{Feedback noise}
\label{subsec:feedback_loop_noise}

\begin{figure*}[htb]
\includegraphics[width=1.7\columnwidth]{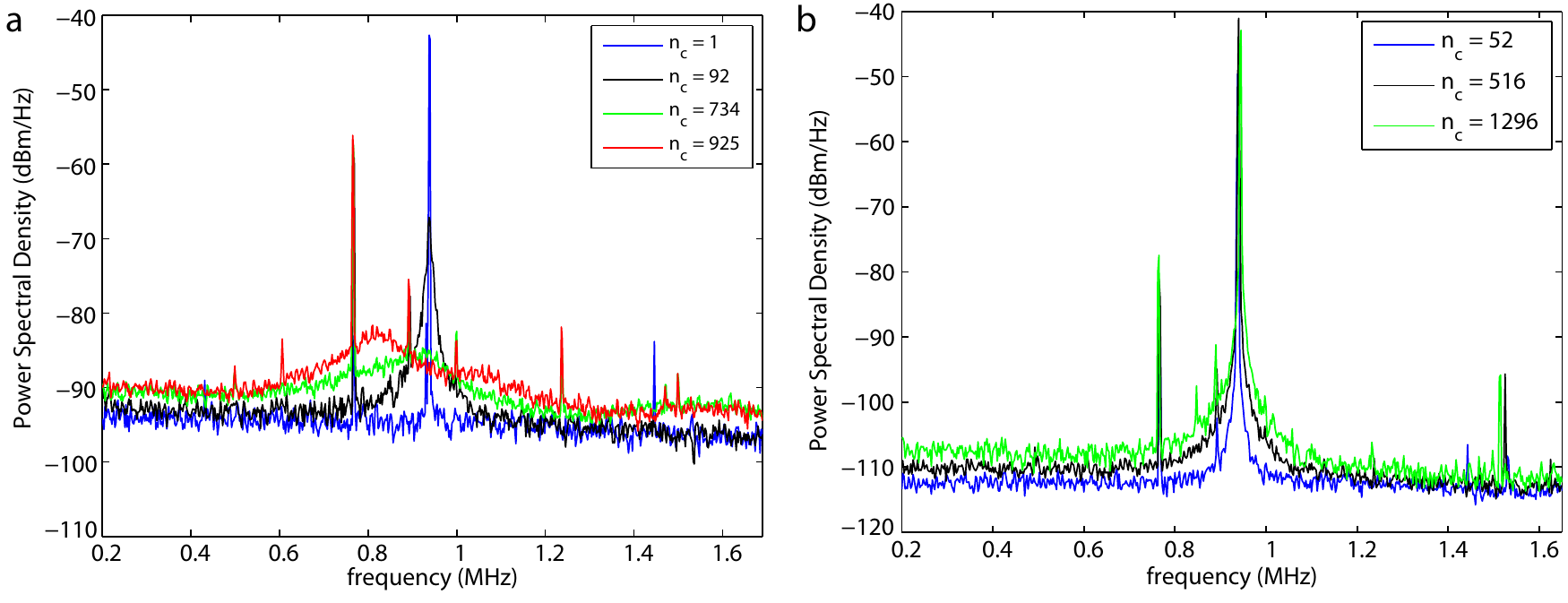}
\caption[]{\textbf{a}, The wide-span noise spectra for a series of experimental powers with the feedback loop engaged (closed loop) and the laser tuned on-resonance with the optical cavity ($\Delta=0, \thetaH=\pi/2$). \textbf{b},  The wide-span noise spectra for a series of experimental powers with the feedback loop disengaged (open loop) and the laser tuned on resonance with the optical cavity ($\Delta=0, \thetaH=\pi/2$). The noise floor at the lowest photon number shown is due to vacuum noise and its slight downward slope is from the detector's response. Note that the discrepancy between the absolute noise levels between the two plots is because the data were taken at different detector gain and local oscillator powers.}\label{SI_WideSpectra_Master}
\end{figure*}

\begin{figure}[htb]
\includegraphics[width=0.8\columnwidth]{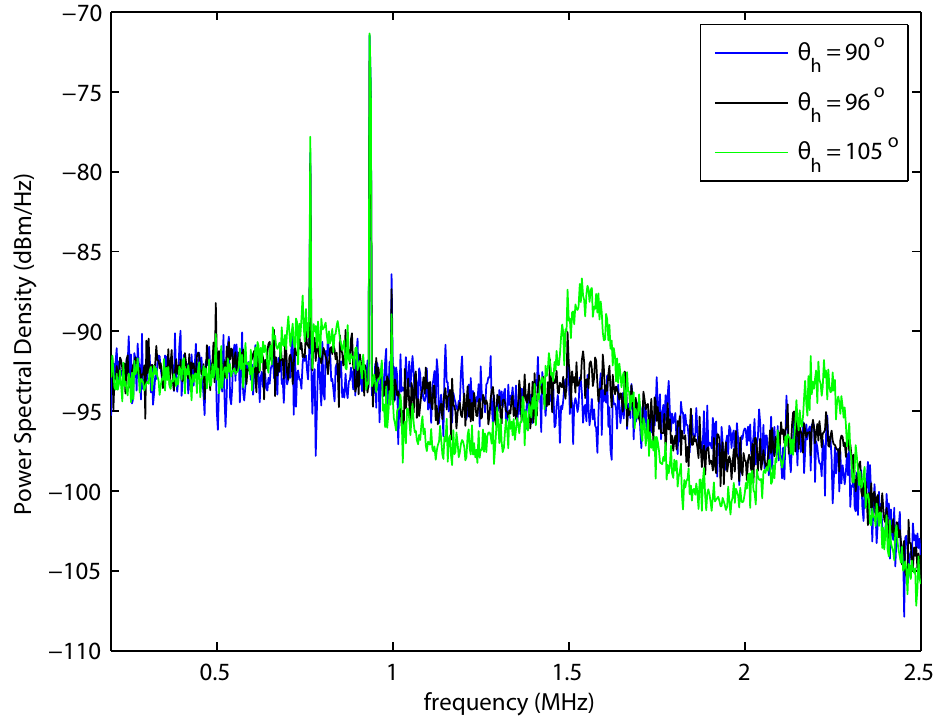}
\caption[]{Wide-span noise spectrum for the feedback loop engaged (closed loop) and the laser far detuned from the laser cavity ($\Delta/\kappa \gg 1$).  Here we plot the spectrum for a series of different homodyne detector phases around $\thetaH = \pi/2$. Measurements were taken at $\ncavO^{\eff}=500$ which corresponds to the number of photons that would be in the cavity to yield the same reflected power at the detector if the laser were on resonance $\Delta =0$.}\label{SI_lockon_offres_noise}
\end{figure}

This section shows the wide span excess noise (above vacuum noise) observed in our measurement setup at high optical power. It also describes how the time delay in the feedback loop from the previous section manifests itself as excess noise in the optical power spectral density at high optical powers. Figure~\ref{SI_WideSpectra_Master}a shows a wide span of the noise power spectral density (NPSD) of our device with the feedback engaged, the probe laser on resonance with the cavity ($\Delta=0$), and the homodyne phase set to the phase quadrature ($\thetaH=\pi/2$). At low optical signal power ($\ncavO \lesssim 100$), the noise floor is set by the vacuum noise of the signal arm and the Lorentzian mechanical response dominates the signal. At higher signal powers ($\ncavO=734$, green) there is $\sim 3$~dB additional broadband noise above vacuum noise which drops off with increasing frequency, but the mechanical response near $940$~kHz still dominates. This broadband noise is also seen in the open loop measurements of Fig.~\ref{SI_WideSpectra_Master}b. A separate measurement indicates that this background noise is not laser intensity noise. Additionally, by measuring the same broadband noise off resonance from the optical cavity ($\Delta/\kappa \gg 1$) we have eliminated the cavity (through thermo-refractive effects) as the source of this excess broadband noise. The noise is therefore likely due to residual optical phase fluctuations in our set-up, possibly due to intrinsic phase noise of the laser or acoustic noise pick up . Note that in principle homodyne detection is insensitive to frequency fluctuations of the laser since the laser is interfered with itself when a common LO is used, however, in practice the two arms of the interferometer are never exactly the same length and the interference is between light emitted by the laser at different times resulting in a conversion of frequency noise to intensity noise which is detected.  Currently this limits our imprecision to about $\nimp = 10^{-4}$ quanta.

In the experiments presented in the main text it is not, however, the broadband excess imprecision noise which ends up limiting the attainable cooling.  as can be seen in Fig.~\ref{SI_WideSpectra_Master}a, at increased laser power ($\ncavO = 925$; red curve) there is a broad noise peak centered near $\sim 800$~kHz which grows and prevents further cooling (or accurate determination of the mode occupancy for that matter). In order to investigate this further we measure the NPSD with the laser off-resonance ($\Delta/\kappa \gg 1$) and the feedback engaged in Fig.~\ref{SI_lockon_offres_noise}. Note that because the data was taken off-resonance, there is very little intracavity power.  For direct comparison then, we have taken the data at $\ncavO^{\eff}=500$, which is the number of photons that would be in the cavity for the same reflected optical power as with the laser directly on resonance ($\Delta = 0$). The feedback loop produces excess noise ``humps'' at approximately $800$~kHz, $1.5$~MHz, and $2.2$~MHz frequencies. The ``humps'' in Fig.~\ref{SI_lockon_offres_noise} correspond in frequency to where the phase of the entire feedback loop (see Fig.~\ref{SI_WideOffRes2}a) is between $0^\text{o}$ and $180^\text{o}$ modulo $360^\text{o}$, which corresponds to the regions where the feedback loop has positive gain, amplifying the measured noise rather than damping it.  This limits the bandwidth over which we can effectively damp and cool the mechanical motion to about $250$~kHz, which for the present device limits the attainable cooling level.   

\end{document}